**Chiral Plasmonic Hydrogen Sensors**

Marcus Matuschek, Dhruv Pratap Singh, Hyeon-Ho Jeong, Maxim Nesterov, Thomas Weiss, Peer Fischer*, Frank Neubrech*, and Na Liu*

M. Matuschek, Dr. D. P. Singh, H.-H. Jeong, Prof. P. Fischer*, Prof. N. Liu*
Max Planck Institute for Intelligent Systems, Heisenbergstr. 3, 70569 Stuttgart, Germany

Dr. M. Nesterov, Jun.-Prof. T. Weiss
4th Physics Institute and Research Center SCoPE, University of Stuttgart, Pfaffenwaldring 57, 70569 Stuttgart, Germany

Prof. P. Fischer*
Institute for Physical Chemistry, University of Stuttgart, Pfaffenwaldring 55, 70569 Stuttgart, Germany
E-mail: fischer@is.mpg.de

M. Matuschek, Dr. F. Neubrech*, Prof. N. Liu*
Kirchhoff-Institute for Physics, Im Neuenheimer Feld 227, 69120 Heidelberg, Germany
E-mail: neubrech@kip.uni-heidelberg.de, laura.liu@is.mpg.de

In this Article, we demonstrate a chiral plasmonic hydrogen sensing platform using palladium-based nanohelices. Such three-dimensional chiral nanostructures fabricated by nano-glancing angle deposition exhibit strong circular dichroism both experimentally and theoretically. The chiroptical properties of the palladium nanohelices are altered upon hydrogen uptake and sensitively depend on the hydrogen concentration. Such properties are well suited for remote and spark-free hydrogen sensing in the flammable range. Hysteresis is reduced, when an increasing amount of gold is utilized in the palladium-gold hybrid helices. As a result, the linearity of the circular dichroism in response to hydrogen is significantly improved. Our chiral plasmonic sensor scheme is of potential interest for hydrogen sensing applications, where good linearity and high sensitivity are required.

## 1. Introduction

Hydrogen is a key energy carrier of the future,[1, 2] for example for fuel cell powered vehicles[3-6] and efficient energy storage devices.[7] Due to its high energy storage capability and light weight, hydrogen is an ideal replacement for fossil fuels such as coal, oil, and natural gases to power combustion engines, turbines, *etc*. Also, hydrogen represents one of the promising alternatives to lithium ion batteries as a mobile energy supply. Nevertheless, the flammability of hydrogen (4-75 vol% in air)[7] together with its ignition at very low energy input brings about critical safety concerns due to potential explosion risks. Therefore, fast and accurate hydrogen sensors with high sensitivity are greatly desirable for versatile real-life applications. To this end, different hydrogen sensors have been implemented based on electrochemical,[8-11] mechanical,[12, 13] catalytic,[14, 15] resistance[16-20] as well as optical schemes.[21-26] In particular, hydrogen sensors can be designed for different application needs and working conditions.[27, 28] Among them, optical sensors offer a great solution to remote readout and are generally safer than others, eliminating spark generation upon hydrogen exposure in the flammable range.

Optical sensors based on plasmonic nanostructures have attracted tremendous attention due to their high sensitivity and noninvasive nature. Illuminated by light at wavelengths much larger than the dimensions of metal nanoparticles, collective oscillations of conduction electrons, the so-called localized surface plasmon resonances (LSPRs) can be excited. The resonance wavelengths, which are commonly used as the readout quantity in LSPP-based sensors, are readily tunable through manipulating the material, size, shape, and local environment of the nanoparticles.[29-31] For instance, resonance position shifts[24] or complete disappearance[32] of the optical responses can be achieved, if hydrogen-responsive plasmonic nanostructures are used. Such optical response changes can be monitored in real time with linear or unpolarized light in reflection, [24] transmission,[23] or interferometric measurements.[33] The resonance shifts are also accompanied with intensity changes at the monitored wavelength positions.[24] Thus, the sensor readout can be conveniently obtained with monochromatic radiation (*e.g.*, a small laser diode), enabling a straightforward integration of the LSPP concept for miniaturized and portable sensor systems.[34]

Previously, plasmonic hydrogen sensors based on phase-transition materials including palladium (Pd), [23, 24, 26] magnesium (Mg)[35], yttrium (Y)[32] as well as metal alloys [22, 25, 36] have been attempted. Among these materials, Pd undergoes a fast and reversible phase transition from a metal to a metal hydride, when exposed to hydrogen. This phase transition is highly dependent on hydrogen concentrations, making Pd ideally suitable for hydrogen sensing.[37, 38] During hydrogen exposure, there exist three different states, *i.e.*, the $\alpha$-phase, $\alpha + \beta$-phase and $\beta$-phase. In the $\alpha$-phase, hydrogen forms a solid solution in the material,[21] while in the $\beta$-phase, hydrogen binds to the Pd atoms, and due to an energy barrier created by the lattice strain, hysteretic effects are observed.[28] The $\alpha + \beta$-phase is a mixed phase. Furthermore, the response time, which is defined as the time to reach 90% of the signal change, peaks in the region between the $\alpha$- phase and the $\beta$-phase.[36]

Hysteresis and long response time of pure Pd nanostructures hamper their applications as hydrogen sensors. Hysteresis affects the sensor performance in two regards. First, a certain hydrogen concentration may correspond to different readout signals (*e.g*., the resonance intensities) during hydrogenation and dehydrogenation cycles. This leads to undefined states, which require precise knowledge of the sensor's history to determine the correct hydrogen concentration. Second, only a limited concentration interval, in which the measured signal (*e.g.*, the resonance wavelength) scales linearly with the concentration, is available. To solve these problems, alloying Pd with other metals such as Au,[22, 25, 36, 39] Ag[36, 40] or Ni[18, 19] has been applied and achieved great success in suppressing hysteresis, while simultaneously reducing the response time.

Compared to standard extinction, reflection, and scattering spectroscopies, circular dichroism (CD) spectroscopy has been proven to allow for more sensitive determination of LSPR shifts in response to the refractive index changes in the particle surrounding.[41-43] Plasmonic chirality results from a theme of handedness from a chiral plasmonic structure when interacting with circularly polarized light. Different top-down[41, 44] and bottom-up[45-48] approaches can be used to prepare a variety of chiral plasmonic structures. However, fabrication of three-dimensional (3D) chiral structures at visible frequencies still remains challenging. Standard nanofabrication methods, such as electron beam lithography[49] and direct laser writing[50, 51] are limited in their capabilities to produce plasmonic structures in large area at low cost and to achieve nanostructures with high-quality metallization, respectively. In contrast, nano-glancing angle deposition (NanoGLAD)[42, 52-54] allows for wafer-scale growth of complex 3D nanostructures, for instance*,* hooks, helices, and zig-zags,[42] with great material flexibility and high geometrical resolution.

In this Article, we demonstrate chiral plasmonic sensors, comprising large-area low-cost

Pd-based nanohelices fabricated by NanoGLAD for hydrogen detection. We show that CD spectroscopy enables hydrogen sensing with an increased sensitivity in comparison to conventional extinction spectroscopy. In particular, using Pd-Au hybrid helices, the response time, linearity, and sensing accuracy can be significantly improved, rendering such nanostructures especially interesting for practical sensing applications.

## 2. Experiments and Discussion

### 2.1 Sample fabrication

We fabricated a series of nanohelices using NanoGLAD,[55] which is a wafer-scale 3D nanofabrication technique that combines block copolymer micelle nanolithography (BCML)[56] with glancing angle deposition (GLAD)[57] as shown in **Figure 1**. This technique allows for precise control over both the shape and material composition of 3D nanostructures. The former was used to adjust the chiral shape factors,[58] while the latter was used to fabricate the helices and engineer the material composition for optimizing the sensing performance.[59] A quasi-hexagonal array of Au nanoseeds was first patterned using BCML with desired particle diameter and spacing (10 nm and 100 nm, respectively) on a quartz substrate (see Figure 1a). Then, the nanohelices were grown on the seeds in a GLAD system based on co-deposition from dual electron-beam evaporators at a vapor flux angle of 87º at 90 K substrate temperature with a base pressure of $1\times10^{-6}$ mbar (see Figure 1b). We adopted a single-turn helical shape, featuring a wire radius of 13 nm, a helix radius of 25 nm, and a pitch of 130 nm as depicted in Figures 1c-1d. Note that, as such a high density of plasmonic helices gives rise to strong extinction (*e.g*., over 2 optical density for 1.5 turns) that hinders the sensing accuracy, we deliberately chose a single-turn helix geometry for achieving strong chiroptical responses, while reducing the optical density close to 1 (see **Figure 2**a). For fabrication of Pd-Au hybrid nanohelices, the crucial parameter was the deposition ratio between Pd and Au, which was determined using two independent quartz crystal microbalances of the individual evaporators and was kept constant over the entire deposition processes. As a result, we were able to grow a series of Pd-Au nanohelices with different Au contents ranging from 0 atom% to 42 atom%, while maintaining their overall size (see **Figure 4**).

**2.2 CD and extinction spectral measurements**

CD is defined as the differential extinction of left- (LCP) and right-handed circularly polarized (RCP) light, when interacting with a chiral object. The extinction spectra of the Pd nanohelices residing on a quartz substrate were measured using a CD spectrometer (Jasco J-1500). A pinhole of 7 mm diameter was placed in front of the sample. A bare quartz substrate was used as a measurement reference. A representative CD spectrum is shown in Figure 2a (black line). Due to the chiral nature of the plasmonic helices, a strong CD response (dip-peak-feature) up to 1.7 deg is observed at 250 nm, which corresponds to the resonance position of the Pd nanohelices (blue line in Figure 2a).

CD and extinction spectra were also simulated using COMSOL Multiphysics for the single-turn Pd helices (pitch 130 nm, wire radius 14 nm, helix radius 30 nm, dielectric data taken from *Johnson and Christy*[60]). The Pd helices were arranged in a hexagonal lattice with a lattice constant of 90 nm. In Figure 2b, the simulated CD and averaged extinction (LCP and RCP) spectra are presented. The experimental and simulated CD spectra show an overall good agreement. Due to approximately 5-10% variations in the structural parameters of the nanohelices grown by nanoGLAD (for example, see Fig. S1, supporting information),[42] weaker signal strengths are observed for both the experimental CD and extinction spectra, when compared to their simulated counterparts. Numerical parameter studies reveal that even small structural changes, for instance, a 2 nm increase of the wire radius, can result in a spectral shift of 30 nm (see **Figure S2,** supporting information). This demonstrates a crucial dependence of the chiral response on the exact geometrical dimensions. A change of the helix radius leads to a similar behavior, whereas a slightly modified periodicity can give rise to a shift of the CD dip (see Figure S2**,** supporting information).

**2.3 Plasmonic Pd nanohelices as hydrogen sensors**

**Figure 3**a presents the working principle of our hydrogen sensor. The Pd nanohelices are illuminated by circularly polarized light during hydrogen exposure. The CD and extinction spectra are altered in dependence on hydrogen concentrations in real time. The acquired spectra during hydrogenation are shown in Figure 3b for the CD (top) and the averaged extinction (bottom), respectively. For both cases, the plasmonic responses gradually decrease, when increasing the hydrogen concentration. The total CD signal change is approximately 50%, whereas the extinction signal change is only 10%. This reveals a high sensitivity of the chiral response upon hydrogen concentration changes.

In the experiment, the hydrogen concentration in nitrogen was varied in steps of 0.1% from 0% to 2.5% and then back to 0% at room temperature, while the flowrate was kept constant at 1 l/min. For the hydrogenation process, three different regimes can be identified. This is depicted in the upper panel of Figure 3c, where the signal strengths of CD at its minimum (270 nm, black) and maximum (660 nm, red) as well as extinction at its maximum (270 nm) normalized to their respective initial values are shown in dependence on hydrogenation concentrations. The filled squares in the upper and lower panels trace the CD and extinction signal evolutions with increasing hydrogen concentration, respectively. First, from 0% to 1.6%, the changes in CD are below 30 mdeg per 0.1%. From 1.7% to 1.9%, abrupt CD changes are observed ranging from 150 to 450 mdeg per 0.1%. With further concentration increases, the CD changes are below 20 mdeg per 0.1%. These three different regimes correspond to the $\alpha$-phase at low hydrogen concentrations, the mixed $\alpha + \beta$-phase, as well as the $\beta$-phase at high hydrogen concentrations. It is worth mentioning that the evolutions of minimum and maximum CD are nearly identical. The extinction data follow a similar behavior (see lower panel in Figure 3c, filled squares): small extinction changes (<0.001) are found in the $\alpha$- and $\beta$-phases, while relatively bigger changes (0.02-0.04) are present in the transition regime, when the hydrogen concentration varies from 1.7% to 1.9%.

The dehydrogenation process as shown in Figure 3c (empty squares) for CD and extinction measurements exhibits clear hysteresis, resulting from the induced lattice strains due to hydrogen incorporation.[28] More specifically, during hydrogenation the phase transition takes place in a range from 1.7% to 1.9%, whereas during dehydrogenation the transition occurs between 0.9% and 0.6% of hydrogen. Such hysteresis behavior makes pure Pd nanostructures less favorable for hydrogen sensing applications. When examining the sensor response time, in the $\alpha$- or $\beta$-phase it is shorter than 120 s. However, in the mixed phase from 1.7% to 1.9%, the response time increases to 40 minutes (see **Figure S3**, supporting information). The long response time also hampers the use of pure Pd nanostructures for practicable sensing applications.

**2.4 Plasmonic Pd-Au hybrid nanohelices**

One elegant solution to reduce hysteresis is to alloy Pd with other metals,[22, 27] for instance nickel,[18, 19] Au[21, 22, 39] or silver,[36, 40]. In order to improve our sensor performance, we fabricated Pd-Au hybrid nanohelices with different Au contents (5, 23, and 42 atom%) using NanoGLAD. Au was the material of choice due to its chemical stability.[21, 22, 39] The dimensions of the hybrid helical structure are similar to those of the pure Pd structures as demonstrated by dark-field transmission electron microscopy (TEM). Energy-dispersive X-ray spectroscopy (EDX) was performed to examine the exact contents and distributions of both materials within the structures. To carry out the TEM measurements, helices were removed from the substrates in an ultrasonic bath (Bandelin Sonorex Super RK 514 BH, 15 minutes[42, 54]). Afterwards, the solution containing free nanohelices was dropped onto a TEM grid and dried for two hours.

Representative TEM images of the nanohelices are shown in the first row of Figure 4a. The single-turn helices show an increase of the wire radius towards the head of the structure due to crystallization. Samples with different material compositions possess similar sizes and shapes. This elucidates the robustness of the NanoGLAD technique. The second and third rows present the atom% as well as Pd and Au elemental mapping in the helices, respectively. It is evident that both materials are distributed homogeneously within the helices, verifying the high quality of the Pd-Au hybrid structures.

CD measurements of the Pd-Au nanohelices were performed during hydrogenation and dehydrogenation. When comparing the CD signals (at its minimum) normalized to the respective initial values, large differences are observed for samples with varied Au contents as shown in Figure 4b. First, hysteresis is significantly reduced with increasing the Au content.[18, 22, 40] Second, the CD signals scale more linearly with the hydrogen concentration, when the Au content is increased. Additionally, the response times are below 20 s (see **Figure 5** and Figure S3 (b)) for all steps during hydrogenation and below 80 s (see Figure 5 and Figure S3 (c)) during dehydrogenation for the samples with 23 atom% and 42 atom%. Although the absolute CD values decrease with increasing the Au content, the CD changes remain sufficiently large enough to sensitively correlate with different hydrogen concentrations. This fact in combination with the improved linearity and short response time makes chiral Pd-Au hybrid nanostructures advantageous for plasmonic hydrogen sensing.

**2.5 Improved sensing performance**

In the following, we focus on the Pd-Au helix sensor with an Au content of 23 atom%, as this configuration possesses a nearly linear relation between the CD signal and hydrogen concentration. Also, its response time is short and the associated CD spectra are distinct. The CD and extinction changes were monitored at a fixed wavelength of 230 nm. Figure 5 shows the time evolutions of the CD and extinction signals during hydrogenation. When the hydrogen concentration is increased to 1.6%, the CD signal is reduced by ~40%, while the change of the extinction signal is only 10% (see Figure 5b). In addition, the signal to noise ratio is substantially higher for the CD measurements. This effect becomes more evident in the time range from 60 to 100 minutes (dashed lines, hydrogen concentration from 1.2% to 1.6% and then to 1.1%). In this time interval, the CD signals at ten different hydrogen concentrations (five for concentration increases and five for concentration decreases), can be easily distinguished. In contrast, the extinction results show a much lower sensitivity and detailed concentration dependence cannot be clearly resolved. This comparison nicely demonstrates the superior performance of CD spectroscopy for hydrogen detection using the chiral Pd-Au hybrid nanohelices. It is noteworthy that the response times are below 20 s for all steps during hydrogenation (see Fig. S3, supporting information), which are slightly longer than the response times of the Pd-based optical hydrogen sensor reported by Wadell, *et al*.[22] There is still plenty of room to reducing the response time, for example, by optimizing the crystalline quality and the alloying mixture so that response times below 1 s are achievable.

## 3. Conclusion

In conclusion, we have presented a novel plasmonic hydrogen sensing scheme using 3D chiral nanohelices. The optical readout has been carried out using CD spectroscopy for hydrogen detection with sensitivity much higher than the standard extinction spectroscopy. The signal changes upon hydrogen uptake by the nanohelices are clearly distinguishable in the CD spectra over a large spectral range (ultraviolet to near infrared), whereas their corresponding extinction spectra show only small changes with low signal to noise ratios, which hamper hydrogen detection at low concentrations with minute variations. Importantly, the response times and hysteresis of the CD spectral changes at different hydrogen concentrations have been significantly suppressed by adjusting the amount of Au in the Pd matrix. Our work suggests a novel solution to producing large-area low-cost optical plasmonic hydrogen sensors, which offer good linearity and high sensitivity for practical applications in addition to the versatile advantages of optical sensors over electrochemical, mechanical, catalytic, and resistance-based detection schemes.

**Supporting Information**

Supporting Information is available from the Wiley Online Library or from the author.

**Acknowledgements**

We thank K. Hahn (MPI FKF) for performing TEM and EDX measurements as well as J. Spatz for SEM access. This project was supported by the Sofja Kovalevskaja grant from the Alexander von Humboldt-Foundation, the Marie Curie CIG grant, and the European Research Council (ERC Dynamic Nano) grant, as well as the ERC grant Chiral MicroBots.

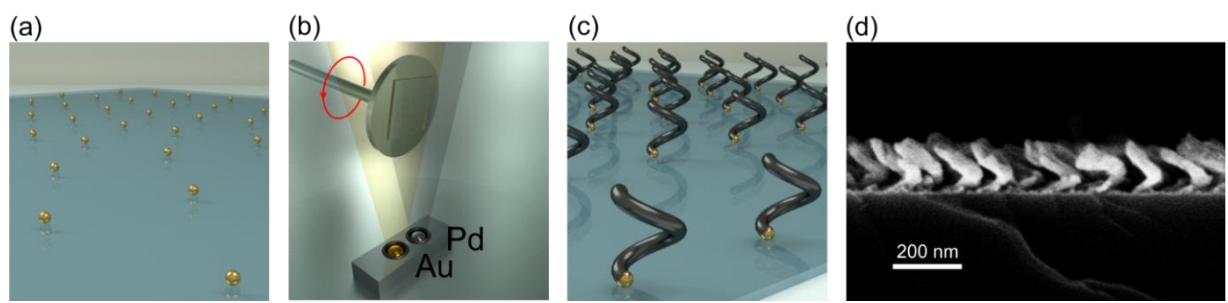

**Figure 1: Fabrication Process of Creating Nanohelices.** (a) Illustration of gold (Au) nanoseeds with an average diameter of 10 nm arranged in a hexagonal lattice on a quartz substrate fabricated by micellar nanolithography. (b) NanoGLAD enables the fabrication of 3D complex nanostructures by tilted angle evaporation and rotation of the substrate. Au and palladium (Pd) can be evaporated simultaneously. (c) Illustration of the periodically arranged Pd nanohelices. (d) Electron micrograph of the as-fabricated Pd nanohelices (cross-section) with a pitch of 130 nm.

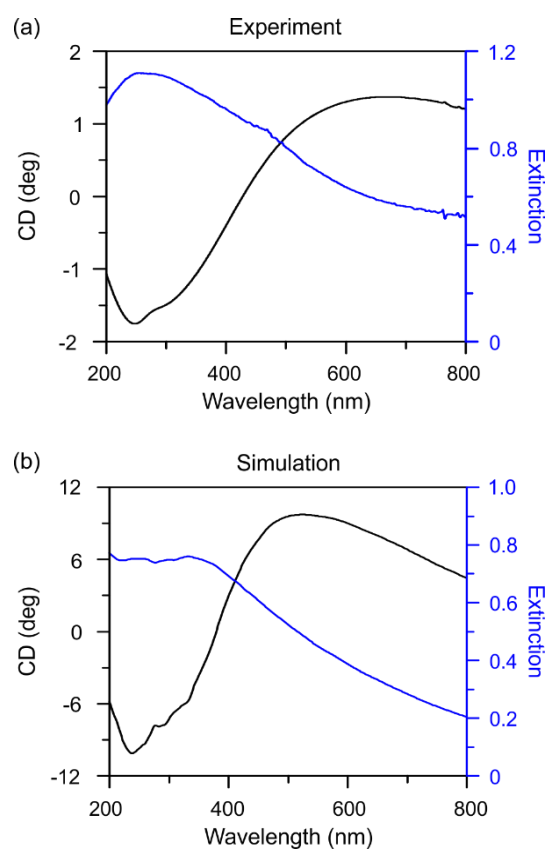

**Figure 2: Experimental and Simulated Optical Spectra.** (a) Measured CD (black) and averaged extinction (blue) spectra of the Pd nanohelices. (b) Simulated CD (black) and averaged extinction (blue) spectra of the Pd nanohelices. Experimental and simulated data are in a good qualitative agreement.

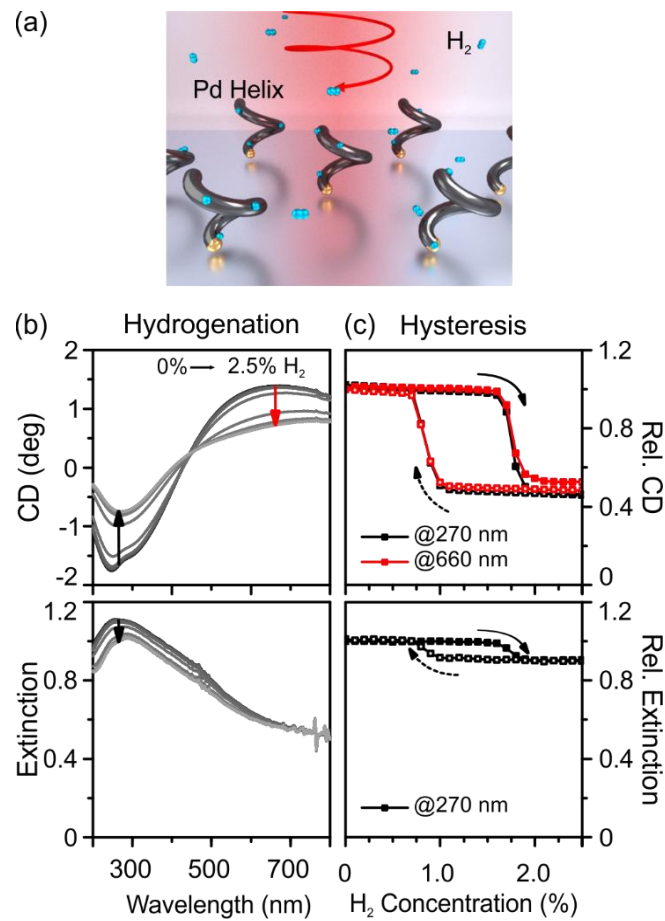

**Figure 3: Sensor Performance during Hydrogenation and Dehydrogenation.** (a) Schematic of the chiral plasmonic hydrogen sensor. Pd nanohelices are illuminated by circularly polarized light, when exposed to hydrogen. (b) CD (upper panel) and extinction (lower panel) spectra during hydrogenation (0% to 2.5% of hydrogen in nitrogen in steps of 0.1%) of the Pd nanohelices. The CD and extinction signals both decrease with increasing hydrogen concentrations. Arrows indicate the signal evolutions during hydrogen uptake. (c) Evolutions of CD (upper panel) and extinction (lower panel) signals normalized to their respective initial values at 0% of hydrogen at fixed wavelengths of 270 nm and 660 nm, respectively, which are also indicated by arrows in (b). Hydrogenation (filled squares) and dehydrogenation (open squares) take place at different hydrogen concentrations, leading to large hysteresis.

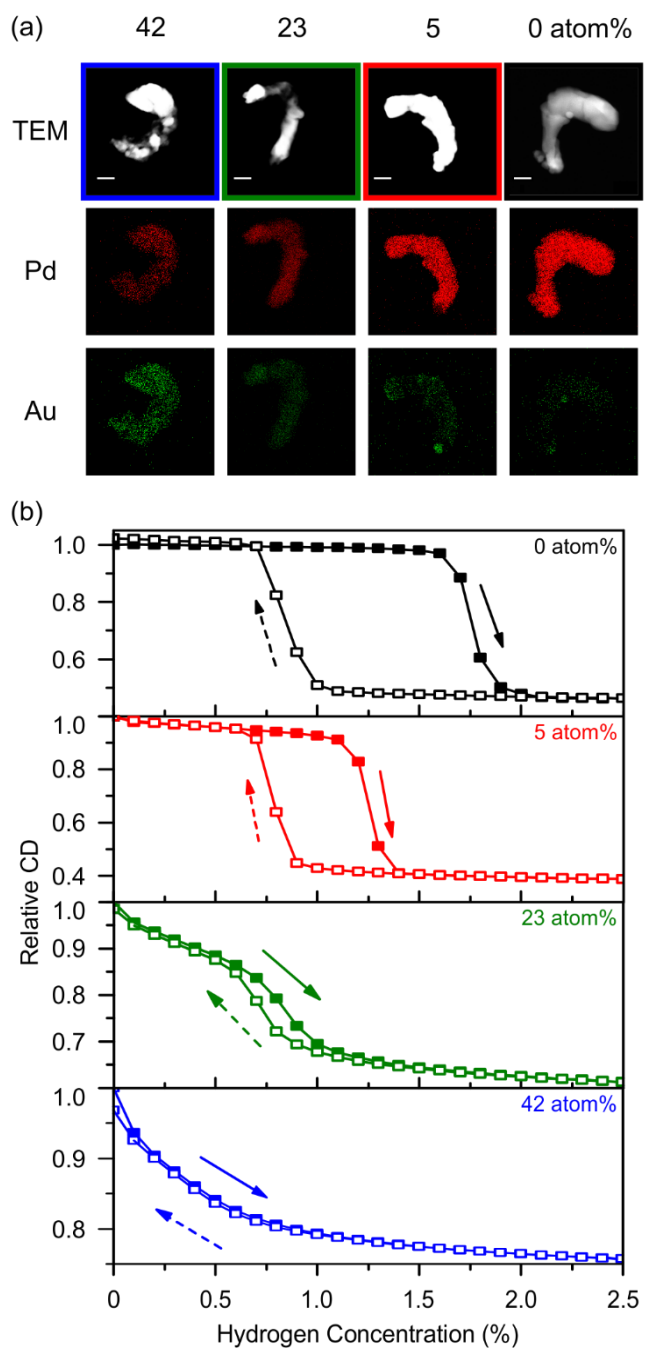

**Figure 4: Elemental Mapping and Hysteresis of Pd-Au Nanohelices with Different Au Contents:** (a) Dark-field TEM and elemental mapping of the Pd-Au nanohelices with different Au contents as given at the top line. Scale bar: 20 nm. All the samples form hybrids with homogeneously spread materials. (b) CD signal evolutions in dependence on hydrogen concentrations traced at the respective minimum CD wavelengths (270 nm, 215 nm, 220 nm and 219 nm) from the samples shown in (a). With increasing the Au content, hysteresis is reduced at the cost of a decreased relative change. Hydrogenation (dehydrogenation) is indicated by a solid (dashed) arrow.

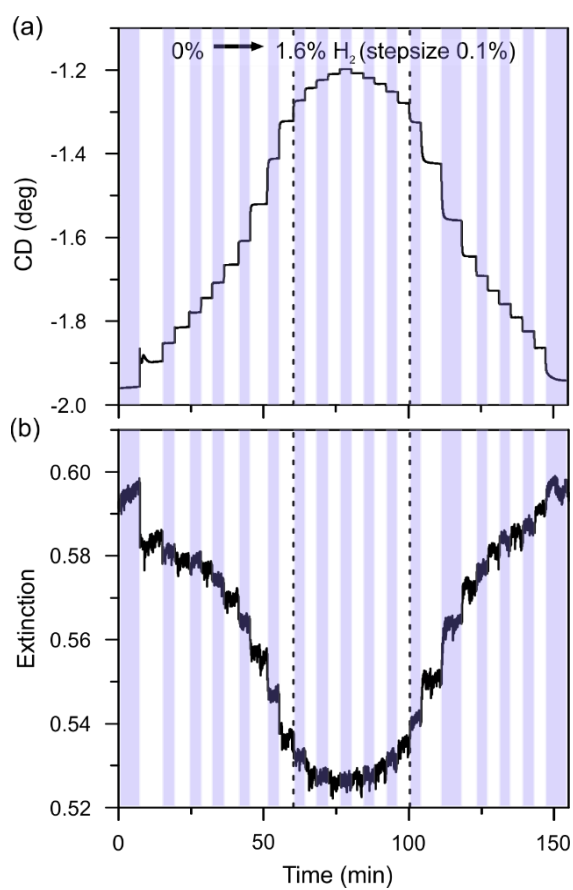

**Figure 5: Optimized Hydrogen Sensor (Au content of 23 atom%).** Time evolutions of (a) CD and (b) extinction signals traced at 230 nm (minimum CD). The hydrogen concentration is changed from 0% to 1.6% (hydrogenation) and 1.6% to 0% (dehydrogenation) in steps of 0.1% (indicated by blue areas). All concentrations can be easily distinguished in the CD measurements, whereas much less distinct steps are observed from the extinction results.

**Chiral plasmonic Pd-based nanohelices for hydrogen sensing.** To reduce hysteresis and improve the linearity during hydrogenation/dehydrogenation, Pd-Au hybrid nanohelices are investigated. Strong CD responses with excellent signal-to-noise ratios are achieved. Such chiral plasmonic sensors are of great interest for hydrogen sensing applications, where good linearity and high sensitivity are required.

Keywords: plasmonics, hydrogen, sensors, palladium, NanoGLAD

**Chiral Plasmonic Hydrogen Sensors**

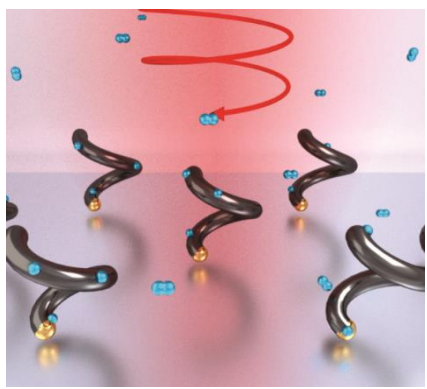

TOC Figure

**Supporting Information**

**Chiral Plasmonic Hydrogen Sensors – Supporting Information**

Marcus Matuschek, Dhruv Pratap Singh, Hyeon-Ho Jeong, Maxim Nesterov, Thomas Weiss, Peer Fischer*, Frank Neubrech*, and Na Liu*

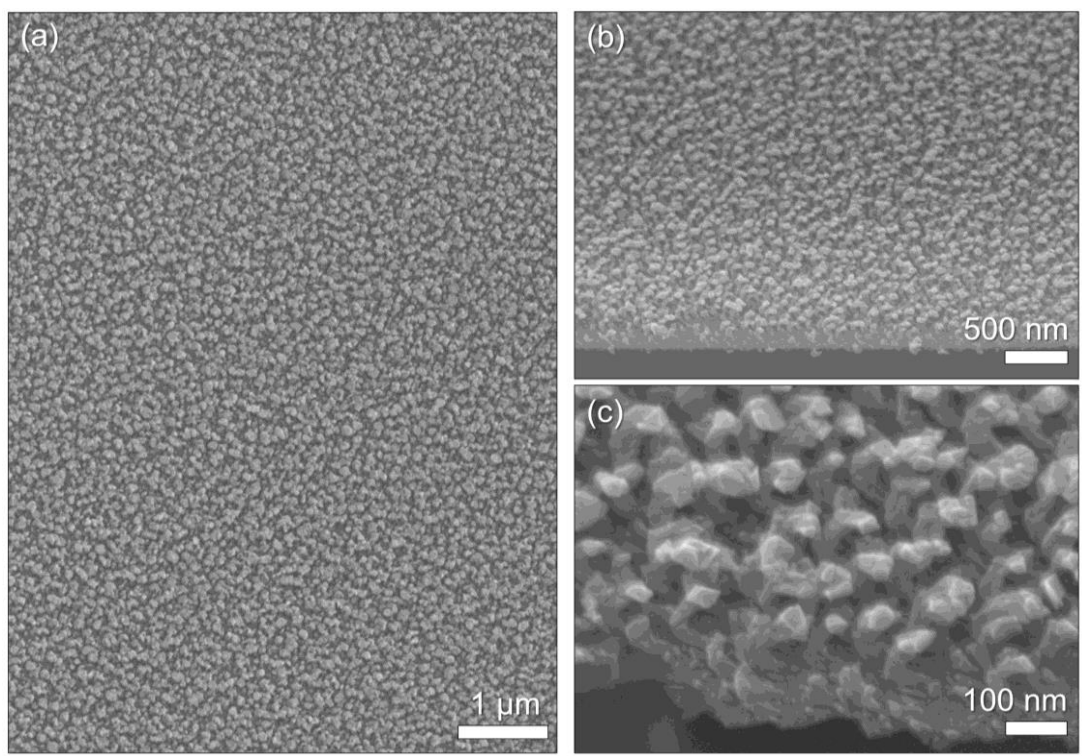

**Figure S1: SEM images of the Pd-Au nanohelices (13% Au).** (a) top-view. (b) and (c) 45° tilted views.

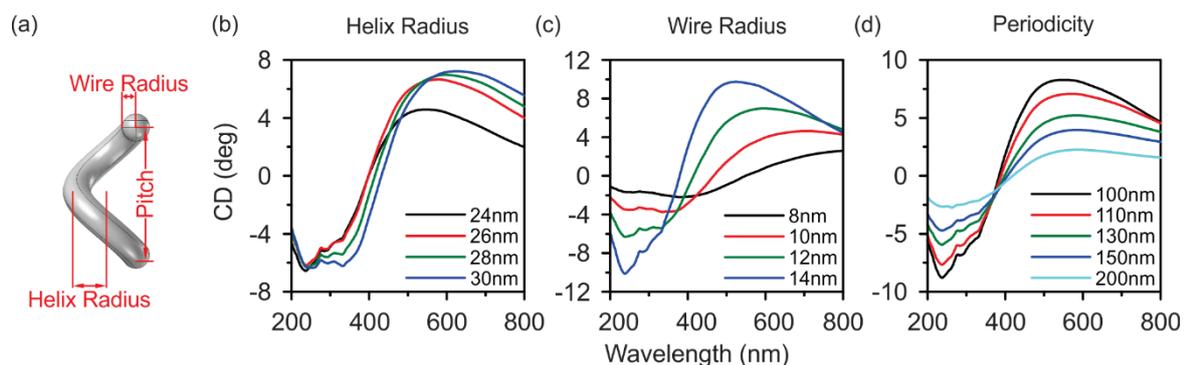

**Figure S2: Simulated CD spectra.** (a) Sketch explaining the different helix paramters. CD spectra for varying helix radii (b), wire radii (c), and periodicities (d). The parameter of interest is varied, while other parameters are fixed (wire radius of 12 nm, helix radius of 30 nm, pitch of 130 nm, and periodicity of 99 nm). The simulations were performed with COMSOL.

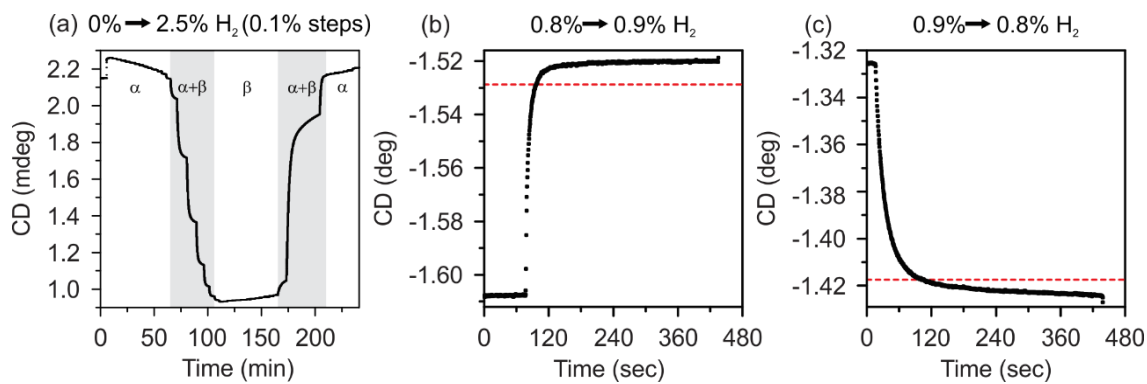

**Figure S3: Maximum CD verses time**. (a) Response time of pure Pd nanohelices strongly depends on the actual phase. In the $\alpha$-phase and $\beta$-phase, the response times are on the order of 120 s, whereas the response time becomes larger than 30 min in the mixed phase ($\alpha+\beta$). The longest response times of a Pd-Au nanohelix sensor with an Au content of 23 atom% during (b) hydrogenation (from 0.8% to 0.9%) and (c) dehydrogenation (from 0.9% to 0.8%). The red line indicates the 90% CD signal change position, which is used to define the response time.